\documentclass[twocolumn,showkeys,showpacs,eqsecnum,nofootinbib,aps]{revtex4}

\def\bea{\begin{eqnarray}}
\def\eea{\end{eqnarray}}

\newcommand{\nn}{\nonumber}
\newcommand{\na}{\nabla}
\def\beq{\begin{equation}}
\def\eeq{\end{equation}}

\begin{document}
\title{Thermodynamics of (1+1) dilatonic black holes in global flat embedding scheme}
\author{Soon-Tae Hong}
\email{soonhong@ewha.ac.kr}
\affiliation{Department of Science
Education, Ewha Womans University, Seoul 120-750 Korea}
\date{\today}%

\begin{abstract}
We study thermodynamics of (1+1) dimensional dilatonic black holes
in global embedding Minkowski space scheme. Exploiting geometrical
entropy correction we construct consistent entropy for the charged
dilatonic black hole. Moreover, (1+1) dilatonic black holes with
higher order terms are shown to possess (3+2) global flat
embedding structures regardless of the details of the lapse
function, and to yield a generic entropy.
\end{abstract}
\pacs{04.70.Dy, 04.62.+v, 04.20.Jb, 11.25.-w}
\keywords{entropy, global flat embedding, IIA string} \maketitle


Since (1+1) dimensional black holes associated with string theory
was proposed~\cite{witten91}, there have been lots of progresses
such as discovery of U-duality between two dimensional dilatonic
black holes~\cite{nappi92,nappi92mod,nappi92plb, gibbons92} and
five dimensional one in the string theory. A thermal Hawking
effect on a curved manifold~\cite{hawk75,beken73} can be looked at
as an Unruh effect~\cite{unr} in a global embedding Minkowski
space (GEMS).  This GEMS approach~\cite{kasner,fronsdal,deser97}
could suggest a unified derivation of thermodynamics for various
curved manifolds~\cite{deser97} and the (5+1) GEMS structure of
(3+1) Schwarzschild black hole solution~\cite{sch} was
obtained~\cite{deser97}.

In this paper we study thermodynamics of (1+1) dilatonic black
holes in the GEMS scheme. Using geometrical entropy correction we
can obtain consistent entropy for a charged dilatonic black hole.
More general (1+1) dilatonic black holes are shown to possess
(3+2) GEMS structures regardless of the details of the lapse
function with higher order terms, and to yield a generic entropy
formula.

We start with two-dimensional dilatonic black
holes~\cite{nappi92,nappi92mod,nappi92plb,gibbons92} associated
with the type IIA string theory and its compactification to five
dimensions whose metric is the product of a three-sphere and an
asymptotically flat two-dimensional geometry.  The ten-dimensional
type IIA superstring solution consists of a solitonic NS 5-brane
wrapping around the compact coordinates, say, $x_{5}$, $x_{i}$
$(i=6,7,8,9)$ and a fundamental string wrapping around $x_{5}$,
and a gravitational wave propagating along $x_{5}$.  In the string
frame, the 10-metric, dilaton and 2-form field $B$ are given
as~\cite{horowitz962,tsey96,mal96,sken98} \bea
ds^{2}&=&-(H_{1}K)^{-1}fdt^{2}\nonumber\\
& &+H_{1}^{-1}K(dx_{5}-(K^{\prime -1}-1)dt)^{2}
\nonumber\\
& &+H_{5}(f^{-1}dr^{2}+r^{2}d\Omega_{3}^{2})+dx_{i}dx^{i},
\nonumber\\
e^{-2\phi}&=&H_{1}H_{5}^{-1},\nonumber\\
B_{05}&=&H_{1}^{\prime -1}-1+\tanh\alpha,\nonumber\\
B_{056789}&=&H_{5}^{\prime -1}-1+\tanh\beta,\nonumber\eea where
$r^{2}=x_{1}^{2}+\cdots+x_{4}^{2}$, $f=1-\frac{r_{0}^{2}}{r^{2}}$
and \bea H_{1}&=&1+\frac{r_{0}^{2}\sinh^{2}\alpha}{r^{2}},~~
H_{5}=1+\frac{r_{0}^{2}\sinh^{2}\beta}{r^{2}},\nonumber\\
K&=&1+\frac{r_{0}^{2}\sinh^{2}\gamma}{r^{2}},~~~ H_{1}^{\prime
-1}=1-\frac{r_{0}^{2}\sinh\alpha\cosh\alpha}{r^{2}H_{1}},\nonumber\\
K^{\prime -1}&=&1-\frac{r_{0}^{2}\sinh\gamma\cosh\gamma}{r^{2}K}.
\nonumber\eea Here $B_{05}$ component of the Neveu-Schwarz 2-form
$B$ is the electric field of fundamental string and $B_{056789}$
is the electric field dual to the magnetic field of the 5-brane
with components $B_{ij}$.  Exploiting dimensional reduction in the $x_{5}$, $x_{i}$
($i=6,7,8,9$) directions in the Einstein
frame~\cite{horowitz962,tsey96}, and then performing an
$T_{5}ST_{6789}ST_{5}$ transformation~\cite{berg95} and an SL(2,R)
coordinate transformation associated with the O(2,2) $T$-duality
group, together with the same set of reverse $S$ and $T$
transformations, one can obtain the five-dimensional black hole
metric \bea
ds^{2}&=&-(H_{1}^{-3}\bar{H}_{5})^{-1/4}K^{-1}fdt^{2}\nonumber\\
& &+(H_{1}^{-3}\bar{H}_{5} )^{-1/4}K(dx_{5}-K^{\prime
-1}-1)dt)^{2}
\nonumber\\
& &+(H_{1}\bar{H}_{5}^{3})^{1/4}(f^{-1}dr^{2}+r^{2}d\Omega_{3}^{2})\nonumber\\
& &+(H_{1}\bar{H}_{5}^{-1})^{1/4}dx_{i}dx^{i},
\label{ds103}\\
e^{-2\phi}&=&\frac{r^{2}}{r_{0}^{2}}+\sinh^{2}\alpha,
\label{dila5} \eea where $\bar{H}_{5}=r_{0}^{2}/r^{2}$.  Next,
performing dimensional reduction in the $x_{5}$, $x_{i}$
($i=6,7,8,9$) directions with $\alpha=\gamma$, one can arrive at
the five-dimensional black hole metric~\cite{teo98} \bea
ds^{2}&=&-\left(1-\frac{r_{0}^{2}}{r^{2}}\right)\left(1+\frac{r_{0}^{2}\sinh^{2}\alpha}{r^{2}}
\right)^{-2}dt^{2}\nonumber\\
& &+\left(\frac{r^{2}}{r_{0}^{2}}-1\right)^{-1}dr^{2}
+r_{0}^{2}d\Omega_{3}^{2},\label{ds533} \eea and the dilaton which
is trivially invariant under the dimensional reduction to yield
the above result (\ref{dila5}).  Here one notes that the metric
(\ref{ds533}) is the product of the two completely decoupled
parts, namely, a three-sphere and an asymptotically flat
two-dimensional geometry which describes the two-dimensional
charged dilatonic black hole. Introducing a new variable $x$  with
$Q=2/r_{0}$ $$
e^{Qx}=2\left(\frac{r^{2}}{r_{0}^{2}}+\sinh^{2}\alpha\right)
(m^{2}-q^{2})^{1/2},$$ where $m$ and $q$ are the mass and charge
of the dilatonic black hole, one can obtain the well-known (1+1)
charged dilatonic black hole~\cite{nappi92,nappi92mod,nappi92plb}
\beq ds^{2}=-N^{2}dt^{2}+N^{-2}dx^{2}, \label{2metric} \eeq where
the lapse function is given as $$N^{2}=1-2m e^{-Qx}+q^{2}e^{-2Q
x}.$$   We can then obtain the horizon $x_{H}$ and $x_{-}$ in
terms of the mass $m$ and the charge $q$:
\bea
e^{Qx_{H}}&=&m+(m^{2}-q^{2})^{1/2},\nonumber\\
e^{Qx_{-}}&=&m-(m^{2}-q^{2})^{1/2}.
\label{eqeq}\eea
By using these relations, we
can rewrite the lapse function as
$$
N^{2}=\left(1-e^{-Q(x-x_{H})}\right)
\left(1-e^{-Q(x-x_{-})}\right).$$

First, we consider the uncharged dilatonic black hole 2-metric
$$ds^{2}=-\left(1-2m e^{-Qx}\right)dt^{2}
+\left(1-2me^{-Qx}\right)^{-1}dx^{2},$$ from which we can
construct  (3+1) dimensional GEMS \bea
z^{0}&=&k_{H}^{-1}\left(1-e^{-Q
(x-x_{H})}\right)^{1/2}\sinh k_{H}t, \nonumber \\
z^{1}&=&k_{H}^{-1}\left(1-e^{-Q (x-x_{H})}\right)^{1/2}
\cosh k_{H}t, \nonumber \\
z^{2}&=&x,\nonumber\\
z^{3}&=&\frac{2}{Q}e^{-Q (x-x_{H})/2}, \label{zzz04} \eea where
the surface gravity is given by $k_{H}=Q/2$.  Using the GEMS
(\ref{zzz04}) and the relation $G_{4}=G_{2}V_{2}$ (details of
which will be discussed later), where $V_{2}$ is a compact volume
$V_{2}=2/Q$ given along $z^{2}$ only, we can then obtain the
desired entropy \beq S=\frac{1}{4G_{4}}\int{\rm d}z^{2}{\rm
d}z^{3}\delta\left(z^{3}-\frac{2}{Q} e^{-Q
(z^{2}-x_{H})/2}\right)=\frac{1}{4G_{2}}, \label{entropy000} \eeq
which is consistent with the previous result
in~\cite{nappi92mod,teo98}.

Second, for a charged dilatonic black hole case associated with
the metric (\ref{2metric}), we can construct a (3+2) GEMS
$ds^{2}=-(dz^{0})^2+(dz^{1})^2+(dz^{2})^2+(dz^{3})^{2}
-(dz^{4})^{2}$ given by the coordinate transformations, \bea
z^{0}&=&k_{H}^{-1}\left(1-e^{-Q (x-x_{H})}\right)^{1/2}
\left(1-e^{-Q (x-x_{-})}\right)^{1/2}\nonumber \\
& &\cdot\sinh k_{H}t, \nonumber \\
z^{1}&=&k_{H}^{-1}\left(1-e^{-Q (x-x_{H})}\right)^{1/2}
\left(1-e^{-Q (x-x_{-})}\right)^{1/2}\nonumber \\
& &\cdot\cosh k_{H}t, \nonumber \\
z^{2}&=&x,\nonumber\\
z^{3}&=&\frac{2}{Q} (1+e^{Q(x_{H}-x_{-})})^{1/2}\sin^{-1}e^{-Q(x-x_{-})/2}
\equiv f(z^{2}),
\nonumber \\
z^{4}&=&\frac{2e^{-3Q(x-x_{H})/2}e^{-Q(x-x_{-})/2}}
{Q(e^{-Q(x-x_{H})}-e^{-Q(x-x_{-})})} \equiv g(z^{2}).
\label{gems04ch}\eea where the surface gravity is given by $$
k_{H}=\frac{Q}{2}(1-e^{-Q(x_{H}-x_{-})}).$$  Here one can also
check that, in the uncharged limit $q\rightarrow 0$, the above
coordinate transformations are exactly reduced to the previous one
(\ref{zzz04}) for the uncharged dilatonic black hole case.
Moreover, one can easily obtain the relation between $z^{3}$ and
$z^{4}$ as follows \bea z^{4}&=&\frac{2e^{3Q(x_{H}-x_{-})/2}}
{Q(e^{Q(x_{H}-x_{-})}-1}\sin^{2}\left[\frac{Qz^{3}}{2(1+e^{Q(x_{H}-x_{-})})^{1/2}
}\right]\nonumber\\
&\equiv& h(z^{3}). \label{z4z3} \eea

In the standard GEMS approach, all the informations for the
entropy come from the areas themselves associated with the event
horizons.  Here the Newton constants $G_{n}$ in the higher
dimensional embeddings are implicitly treated to be the same as the
original $G_{d}$ of the $d$-dimensional black
holes~\cite{deser97}. However, in the (1+1) dilatonic black hole
cases, we could not obtain the areas in terms of the event
horizons due to the delta-function-like behaviors at the event
horizons, which are characteristics of the (1+1) dilatonic black
holes.  As in (\ref{entropy000}), in order to obtain the
entropies, we thus exploit an alternative scheme, where the
entropy informations are extracted from the Newton constants
$G_{n}$ which are now splitted into two factors:
$G_{n}=G_{d}\times V_{n-d}$ with the volumes of the compact
manifolds $V_{n-d}$.  To be more specific, in order to calculate
the entropy for the charged dilatonic black hole, we first
consider a detector on the event horizon at $x=x_{H}$ where the
detector only sees a compact manifold $V_{3}$ along the $z_{3}$
and $z_{4}$ directions, given by \bea V_{3}(x_{H})&=&\int{\rm
d}z^{2}{\rm d}z^{3}{\rm d}z^{4}\delta(z^{3}-f(z^{2}))
\delta(z^{4}-f(z^{3}))\nonumber\\
&=&z^{4}(x_{H}). \nonumber\eea The Newton constant is then given
by $G_{5}=G_{2}V_{3}(x_{H})$ to yield the entropy at $x=x_{H}$
\bea S(x_{H})&=&\frac{1}{4G_{5}}\int{\rm d}z^{2}{\rm d}z^{3}{\rm
d}z^{4}
\delta(z^{3}-f(z^{2}))\delta(z^{4}-h(z^{3}))\nonumber\\
&=&\frac{1}{4G_{2}}. \label{entropy111} \eea Note that, even
though we have used the $x_{H}(q)$ in calculation of the above
entropy $S(x_{H})$, the final result does not contain any
information of the charge $q$ and mass $m$ associated with the
even horizons $x_{H}$ and $x_{-}$,  to yield the same entropy
(\ref{entropy000}) of the uncharged case.

Different from the uncharged case, we have another event horizon
$x=x_{-}$, where we have another compact manifold with volume
$V_{3}(x_{-})=z^{4}(x_{-})$ to yield the modified Newton constant
$\tilde{G}_{5}=G_{2}\tilde{V}_{3}$ with
$\tilde{V}_{3}=V_{3}(x_{H})+V_{3}(x_{-})
=z^{4}(x_{H})+z^{4}(x_{-})$, since the detector at the event
horizon $x=x_{-}$ can see two compact manifolds at $x=x_{-}$ and
$x=x_{H}$.  Moreover, it has been claimed in~\cite{hod99} that the
entropy of a charged black hole should decrease with the absolute
value of the black hole charge.  We can then obtain the entropy
loss due to the existence of the compact manifold at $x=x_{-}$
\bea \delta S&=&\frac{1}{4\tilde{G}_{5}}\int{\rm d}z^{2}{\rm
d}z^{3}{\rm d}z^{4}
\delta(z^{3}-f(z^{2}))\delta(z^{4}-h(z^{3}))\nonumber\\
&=&\frac{1}{4G_{2}}\frac{z^{4}(x_{H})}{z^{4}(x_{H})+z^{4}(x_{-})}.
\nonumber\eea to yield the total entropy $S=S(x_{H})-\delta S$ of
the dilatonic charged black hole as follows \bea
S&=&\frac{1}{4G_{2}}\frac{z^{4}(x_{-})}{z^{4}(x_{-})+z^{4}(x_{-})}\nonumber\\
&=&\frac{1}{4G_{2}}\frac{m+(m^{2}-q^{2})^{1/2}}{2m},
\label{entropyfinal} \eea which is consistent with the previous
result in~\cite{nappi92mod,teo98}.  Note that in the vanishing
charge limit $q\rightarrow 0$, the above entropy is reduced to
that of uncharged case (\ref{entropy000}).  Moreover, without the
$U$-duality transformations discussed above, we can obtain the
consistent entropy (\ref{entropyfinal}) via the GEMS embeddings
and their associated geometrical entropy corrections.

Following the standard procedure in general relativity, one can obtain the 2-acceleration,
the Hawking temperature and the black hole temperature
\bea
a_{2}&=&\frac{Qe^{-Qx}(m-q^{2}e^{-Qx})}
{(1-e^{-2Q (x-x_{H})})^{1/2}(1-e^{-2Q (x-x_{-})})^{1/2}},\nn\\
T_{H}&=&\frac{a_{5}}{2\pi}\nonumber\\
&=&\frac{Q}{4\pi} \frac{1-e^{-Q(x_{H}-x_{-})}}{\left(1-e^{-Q
(x-x_{H})}\right)^{1/2}
\left(1-e^{-Q (x-x_{-})}\right)^{1/2}},\nonumber\\
T&=&NT_{H}=\frac{Q}{4\pi}\left(1-e^{-Q (x_{H}-x_{-})}\right),
\nonumber\eea where we have used the Killing vector
$\xi=\partial_{t}$ on the two-dimensional dilatonic charged black
hole manifold described by $(t,x)$ for the trajectories.  Here
note that the above Hawking temperature $T_{H}$ is also given by
the relation in (\ref{t0g})~\cite{hawk75}.


Next, we consider more general dilatonic black holes associated
with the on-shell action~\cite{nappi92mod,nappi92plb} $$I=\int
d^{2}x\left[-4\na^{a}(e^{-2\phi}\na_{a}\phi)
+e^{-2\phi}(R+2\na^{2}\phi) \right],$$ where the dilaton field is
given by $\phi(x)=\phi_{0}-\frac{1}{2}Qx$ and the 2-metic is given
by (\ref{2metric}) with the lapse function \beq
N^{2}=1+\sum_{n=1}c_{n}e^{-nQx}, \label{lapseg} \eeq where
$c_{1}=-2m$, $c_{2}=q^{2}$ and $c_{n}$ $(n\ge 3)$ are coefficients
of higher order terms.  Note that the lapse function
(\ref{lapseg}) can be rewritten in terms of the event horizons
$x_{n}$ $(n=1,2,...)$ with $x_{1}=x_{H}$ and $x_{n}>x_{n+1}$,
$$N^{2}=\prod_{n=1}(1-e^{-Q(x-x_{n})}).$$  As in the previous case,
we can obtain the surface gravity, the 2-acceleration and the
Hawking temperature in more general form \beq
k_{H}=N\frac{dN}{dx}|_{x=x_{H}},~ a_{2}=\frac{dN}{dx},~
T_{H}=\frac{1}{2\pi }\frac{k_{H}}{N}, \label{t0g}\eeq which are
independent of the dimensionality of the GEMS structures.  We
construct the GEMS embedding solutions for our general (1+1)
dilatonic black hole by making an ansatz of three coordinates
$(z^{0}, z^{1},z^{2})$ in (\ref{zzz0g}) to yield \bea
& &-(dz^0)^2+(dz^1)^2+(dz^2)^2\nonumber\\
& &~~~=ds^{2}-\left(N^{-2}-k_{H}^{-2}\left(\frac{dN}{dx}\right)^{2}-1\right)dx^{2}\nn\\
& &~~~\equiv ds^{2}-(dz^{3})^{2}+(dz^{4})^{2}. \nonumber\eea Here
we have used the fact that the terms in the parenthesis in the
second line can be expressed in terms of difference of two
positive definite terms \beq
N^{-2}-k_{H}^{-2}\left(\frac{dN}{dx}\right)^{2}-1\equiv
F^{2}-G^{2}, \label{twoterms} \eeq where $F$ and $G$ can be read
off from (\ref{zzz0g}).  We can thus obtain the (3+2) dimensional
GEMS $ds^{2}=-(dz^{0})^2+(dz^{1})^2+(dz^{2})^2
+(dz^{3})^2-(dz^{4})^2$ given by the coordinate transformations,
\bea
z^{0}&=&k_{H}^{-1}N\sinh k_{H}t, \nonumber \\
z^{1}&=&k_{H}^{-1}N\cosh k_{H}t, \nonumber \\
z^{2}&=&x,\nonumber\\
z_{3}&=&\int dx F(x)\equiv f(z^{2}),
\nonumber \\
z^{4}&=&\int dx G(x)\equiv g(z^{2}). \label{zzz0g} \eea Note that,
as in (\ref{z4z3}),  $z^{4}$ can be expressed in terms of $z^{3}$:
$z^{4}=g\cdot f^{-1} (z^{3})\equiv h(z^{3})$.

Now we comment on the dimensionality of the GEMS embeddings in the
general dilatonic black holes.  The charge parameter $c_{2}=q^{2}$
introduces one more time-like dimension to yield two time
dimensionalities with (3+2) GEMS structures for the charged
dilatonic black hole.  In the general dilatonic black holes with
$c_{n}$ ($n=1,2,...$), even though we have horizons $x_{n}$ more
than two ones $x_{H}$ and $x_{-}$ of the charged dilatonic black
hole, the GEMS structures are fixed as (3+2) dimensions with no
more increasing dimensionality, since only two positive definite
terms $F^{2}$ and $G^{2}$ are enough to describe the terms of
(\ref{twoterms}) regardless of whatever the lapse function $N^{2}$
has higher order terms with $c_{n}$ $(n=1,2,...)$.

Next, we calculate the entropy for the general (1+1) dilatonic
black hole with higher order terms. As in the charged dilatonic
black hole case, a detector on the event horizon at $x=x_{H}$ only
sees a compact manifold $V_{3}$ along the $z_{3}$ and $z_{4}$
directions to yield the entropy (\ref{entropy111}) at $x=x_{H}$.
However, we have other event horizons $x=x_{n}$ $(n=2,3,...)$
associated with compact manifolds with volumes
$V_{3}(x_{n})=z^{4}(x_{n})$ to yield the Newton constant $$
\tilde{G}_{5}=G_{2}\sum_{n=1}V_{3}(x_{n})=G_{2}\sum_{n=1}z^{4}(x_{n}).
$$ The existences of the compact manifolds at
$x=x_{n}$ $(n=2,3,...)$ thus yield the geometrical entropy
correction originated from $\tilde{G}_{5}$
$$\delta S=\frac{1}{4G_{2}}\frac{z^{4}(x_{H})}{\sum_{n=1}z^{4}(x_{n})}.
$$ so that, together with $S(x_{H})$ which
has the same form as (\ref{entropy111}), we can obtain the total
entropy $S=S(x_{H})-\delta S$ of the general (1+1) dilatonic black
hole \beq
S=\frac{1}{4G_{2}}\frac{\sum_{n=2}z^{4}(x_{n})}{\sum_{n=1}z^{4}(x_{n})}.
\label{entropyfinalg} \eeq

In the charged case with $c_{1}=-2m$, $c_{2}=q^{2}$ and $c_{n}=0$ $(n\ge 3)$,
by exploiting the explicit expression for $z^{4}$
in the GEMS (\ref{gems04ch}) we obtain for the horizons
$x_{1}=x_{H}$ and $x_{2}=x_{-}$ \bea
z^{4}(x_{1})&=&\frac{2e^{Q(x_{1}+x_{2})/2}}{Q(e^{Qx_{1}}-e^{Qx_{2}})},\nonumber\\
z^{4}(x_{2})&=&\frac{2e^{Q(3x_{1}-x_{2})/2}}{Q(e^{Qx_{1}}-e^{Qx_{2}})}.
\label{z4z4} \eea After some algebra with the identities
(\ref{eqeq}), substitution of $z^{4}(x_{1})$ and $z^{4}(x_{2})$ in
(\ref{z4z4}) into the generic entropy formula
(\ref{entropyfinalg}) reproduces the previous result
(\ref{entropyfinal}).  Similarly, for the uncharged case with
$c_{1}=-2m$ and $c_{n}=0$ $(n\ge 2)$, we can easily check that the
entropy (\ref{entropyfinalg}) is reduced to the previous one
(\ref{entropy000}). For more general cases with $c_{1}=-2m$,
$c_{2}=q^{2}$ and nonvanishing $c_{n}$ $(n\ge 3)$, we can find the
expression for $z^{4}$ in the GEMS (\ref{zzz0g}), which is given
by an integral form.  Different from the charged case with
$z^{4}(x_{n})$ $(n=1,2)$ in (\ref{z4z4}), for this general
dilatonic black hole we do not have explicit analytic expressions
for $z^{4}(x_{n})$ at the moment so that we cannot proceed to
evaluate the entropy via the formula (\ref{entropyfinalg}).
However, if the coefficients $c_{n}$ are given explicitly, we can
find $x_{n}$ and $z^{4}(x_{n})$, with which the generic entropy
(\ref{entropyfinalg}) is supposed to yield to all order a result
consistent with that given in \cite{nappi92mod}.


In conclusion, we have investigated the higher dimensional global
flat embeddings of (1+1) (un)charged and general dilatonic black holes.  These
two dimensional dilatonic black holes are shown to be embedded in
the (3+1) and (3+2)-dimensions for the uncharged and
charged two-dimensional dilatonic black holes, respectively.  Moreover, in the general
dilatonic black holes with higher order terms, even though we have horizons $x_{n}$
more than two ones $x_{H}$ and $x_{-}$ of the charged dilatonic black hole,
the GEMS structures have been shown to be fixed as (3+2) with no more increasing
dimensionality.

Different from the uncharged case, in order to obtain the entropy
of the (1+1) charged dilatonic black holes, we have taken into
account all the compact manifold associated with the event
horizons to yield the modified Newton constant.  Exploiting the
geometrical entropy correction originated from the modified Newton
constant, we have obtained the entropy for the charged dilatonic
black holes and even for the general dilatonic black holes.  It is
quite significant to obtain the consistent entropies through the
GEMS embeddings and their associated geometrical entropy
corrections, without getting involved in the $U$-duality
transformations associated with the type IIA string theory.

\noindent {\bf Acknowledgments} The author would like to thank
Prof. Chiara R. Nappi for helpful discussions and worm hospitality
at Princeton University, where a part of this work has been done.
This work was supported by the Ewha Womans University Research
Grant of 2004.


\begin{thebibliography}{99}
\bibitem{witten91} E. Witten, On string theory and black holes, Phys. Rev. D 44 (1991) 314.
\bibitem{nappi92} M.D. McGuigan, C.R. Nappi and S.A. Yost, Charged black holes in two-dimensional
string theory, Nucl. Phys. B 375 (1992) 421.
\bibitem{nappi92mod} C.R. Nappi and A. Pasquinucci, Thermodynamics of two-dimensional black
holes, Mod. Phys. Lett. A 7 (1992) 3337.
\bibitem{nappi92plb} D. Lechtenfeld and C.R. Nappi, Dilaton gravity and no-hair theorem in two dimensions,
Phys. Lett. B 288 (1992) 72.
\bibitem{gibbons92} G.W. Gibbons and M.J. Perry, The physics of 2-D stringy
space-times, Int. J. Mod. Phys. D 1 (1992) 335.
\bibitem{hawk75} S.W. Hawking, Particle creation by black
holes, Commun. Math. Phys. 43 (1975) 199.
\bibitem{beken73} J.D. Bekenstein, Black holes and entropy, Phys. Rev. D 7 (1973) 2333.
\bibitem{unr} W.G. Unruh, Notes on black-hole evaporation, Phys. Rev. D 14 (1976) 870.
\bibitem{deser97} S. Deser and O. Levin, Accelerated detectors and temperature in
(anti)-de Sitter spaces, Class. Quantum Grav. 14 (1997) L163.
\bibitem{kasner}  E. Kasner, Finite representation of the solar
gravitational field in flat space of six dimensions, Am. J. Math.
43 (1921) 130.
\bibitem{fronsdal}  C. Fronsdal, Completion and embedding of the Schwarzschild
solution, Phys. Rev. 116 (1959) 778.
\bibitem{sch}  K. Schwarzschild, \"Uber das gravitationsfeld eines massenpunktes
nach der Einsteinschen theorie, Sitzber. Deut. Akad. Wiss. Berlin,
KI. Math.-Phys. Tech. (1916) 189-196 .
\bibitem{horowitz962} G.T. Horowitz, J.M. Maldacena and A. Strominger,
Nonextremal black hole microstates and U-duality, Phys. Lett. B
383 (1996) 151.
\bibitem{tsey96} A.A. Tseytlin, Extreme dyonic black holes in string theory,
Mod. Phys. Lett. A 11 (1996) 689.
\bibitem{mal96} J.M. Maldacena, {\it Black holes in string theory}, Ph.D
Thesis (Princeton University), hep-th/9607235.
\bibitem{sken98} K. Sfetsos and K. Skenderis, Microscopic derivation of the
Bekenstein-Hawking entropy formula for nonextremal black holes,
Nucl. Phys. B 517 (1998) 179.
\bibitem{berg95} E. Bergshoeff, C. Hull and T. Ortin, Duality in the type II superstring
effective action, Nucl. Phys. B 451 (1995) 547.
\bibitem{teo98} E. Teo, Statistical entropy of charged two-dimensional black
holes, Phys. Lett. B 430 (1998) 57.
\bibitem{hod99} S. Hod, Improved upper bound to the entropy of a charged system,
Phys. Rev. D 61 (2000) 024023.
\end{thebibliography}
\end{document}